\documentclass[conference]{IEEEtran}
\IEEEoverridecommandlockouts

\usepackage{amsmath,amssymb,amsfonts}
\usepackage{bbm}
\usepackage{bm}

\usepackage[table,dvipsnames]{xcolor}  

\usepackage{booktabs}      
\usepackage{multirow}      
\usepackage{array}         
\usepackage{colortbl}      
\usepackage{arydshln}      
\usepackage{threeparttable} 
\usepackage{adjustbox}     

\usepackage{algorithm}     
\usepackage{algorithmic}   

\usepackage{graphicx}      
\usepackage{wrapfig}       

\usepackage{listings}      

\usepackage{tcolorbox}     
\tcbuselibrary{breakable}  

\usepackage{textcomp}      
\usepackage{pifont}        

\usepackage{enumitem}      

\usepackage{cite}          
\usepackage{hyperref}      

\definecolor{light-gray}{gray}{0.92}
\definecolor{light-blue}{rgb}{0.85, 0.91, 0.98}
\definecolor{LightYellow}{rgb}{1.0, 1.0, 0.88}
\definecolor{LightCyan}{rgb}{0.88, 1.0, 1.0}
\definecolor{none}{rgb}{1.0, 1.0, 1.0}

\def\BibTeX{{\rm B\kern-.05em{\sc i\kern-.025em b}\kern-.08em
    T\kern-.1667em\lower.7ex\hbox{E}\kern-.125emX}}
\begin{document}


\title{VFlow: Discovering Optimal Agentic Workflows \\ for Verilog Generation}

\author{
	\IEEEauthorblockN{
		Yangbo Wei $^{1,2}$, 
		Zhen Huang $^{3,2}$, 
		Huang Li   $^{1,2}$, 
		Wei W. Xing$^{4}$,
            Ting-Jung Lin$^{2}$,
		Lei He $^{*2}$ }
    \IEEEauthorblockA{$^1$Shanghai Jiao Tong University, Shanghai, China}
    \IEEEauthorblockA{$^2$ Ningbo Institute of Digital Twin, Eastern Institute of Technology, Ningbo, China}
    \IEEEauthorblockA{$^3$University of Science and Technology of China, Hefei, China}
    \IEEEauthorblockA{$^4$University of Sheffield}
}

\maketitle

\begin{abstract}

Hardware design automation faces challenges in generating high-quality Verilog code efficiently. This paper introduces VFlow, an automated framework that optimizes agentic workflows for Verilog code generation. Unlike traditional approaches relying on fixed prompts or manually designed flows, VFlow treats workflow discovery as a search over graph-structured LLM invocation sequences. It introduces a multi-population cooperative evolution (CEPE-MCTS) algorithm that balances multiple hardware objectives—functional correctness, area, power, timing and token cost—while sharing successful patterns and avoiding repeated failures. Integrated multi-level verification ensures syntactic correctness, functional behavior, and synthesizability. Experiments on VerilogEval and RTLLM2.0 show VFlow improves pass@1 by 20–30\% over prompting baselines and closely matches designer-level area/power. Remarkably, VFlow enables small LLMs to outperform larger models with up to 10.9× ROI, offering a cost-effective solution for RTL design. This work paves the way for intelligent, automated hardware development, advancing LLM applications in EDA.


\end{abstract}

\begin{IEEEkeywords}
Hardware Description Languages, Verilog, Large Language Models, Automated Workflow Optimization, Monte Carlo Tree Search, Digital Circuit Design
\end{IEEEkeywords}

\section{Introduction}

Recent advances in Large Language Models (LLMs) have demonstrated remarkable potential for Verilog code generation. Existing research such as VerilogEval and RTLLM series \cite{liu2023verilogeval, lu2024rtllm} has not only provided standardized benchmarks for evaluating LLMs' ability to generate RTL code but also promoted the automated generation of complex modules ranging from combinational circuits to finite state machines.

Meanwhile, methods such as VeriSeek, RTLCoder, and CodeV \cite{wang2024large,liu2024rtlcoder,zhu2025codev} have enhanced models' adaptability to HDL syntax structures and parallel characteristics through reinforcement learning, structured fine-tuning, and reinforcement feedback mechanisms, respectively. Although these methods have further improved the accuracy and generalization capability of Verilog generation, \textbf{most rely on limited prompt engineering or manually designed multi-stage workflows}, making it difficult to efficiently explore the design space or integrate with hardware verification processes. In contrast, agent-driven automated workflow optimization frameworks (such as AFlow) \cite{zhang2024aflow} represent LLM-driven processes as graph structures and apply MCTS to achieve task-generic optimization. However, these approaches \textbf{have not yet fully incorporated the structured requirements of hardware design and synthesis verification paths}.

\begin{figure}[htbp]

\centering
\includegraphics[width=0.95\linewidth]{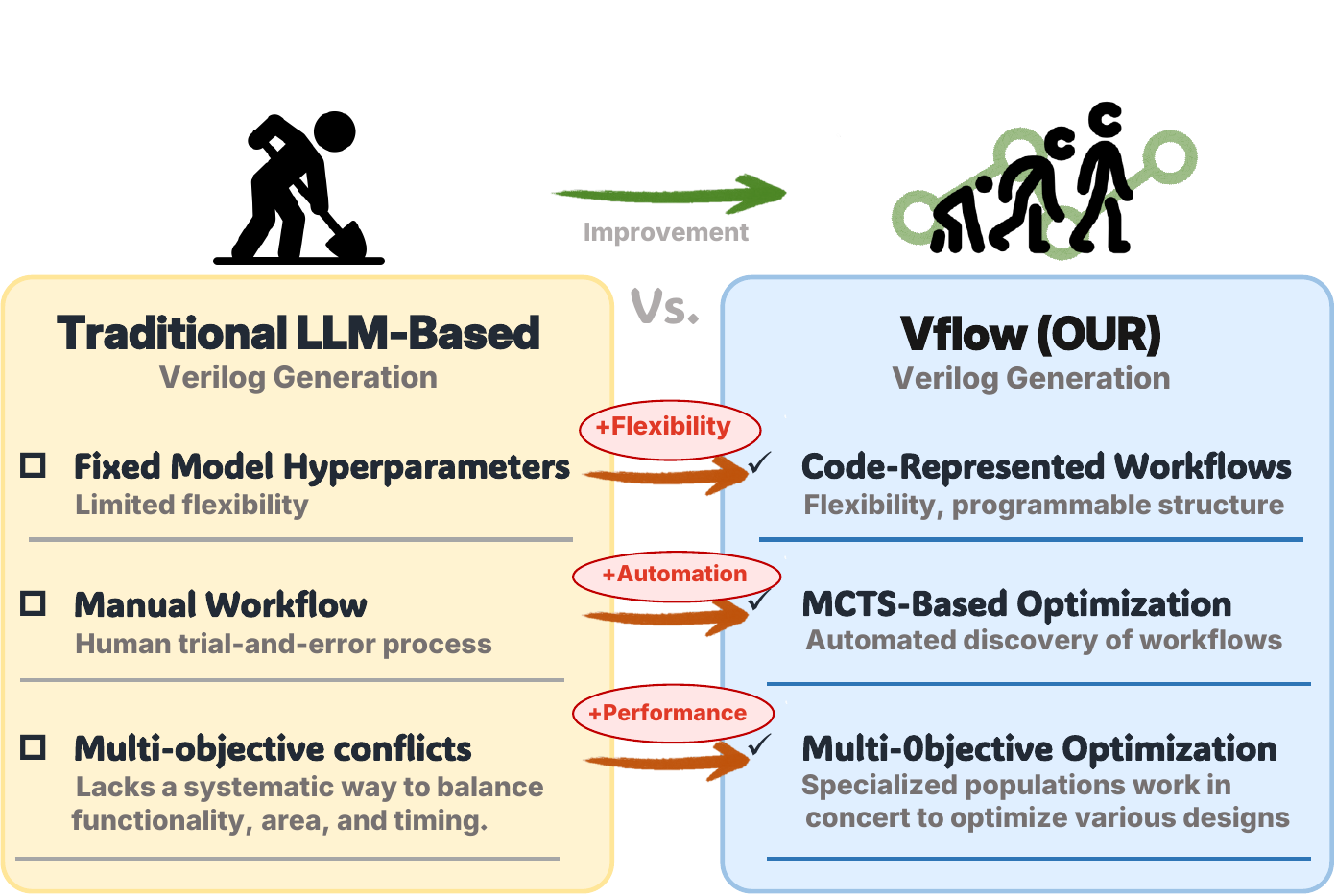}
\caption{Comparison of Vflow with traditional methods.}
\label{fig:motiv}
\vspace{-0.3cm}
\end{figure}

To address these challenges, this paper proposes \textbf{VFlow}, an automated agent workflow optimization framework specifically designed for Verilog code generation tasks. Unlike traditional methods as shown in Figure~\ref{fig:motiv}, VFlow reformulates the workflow optimization problem as a search task on a graph of LLM nodes with code-based representations, employing a \textbf{Cooperative Evolution with Past Experience MCTS (CEPE-MCTS) algorithm}. This approach utilizes parallel multi-population groups focusing on different optimization objectives (such as functionality, area, timing, token cost and balance), combined with fragment migration mechanisms and global failure experience sharing strategies, effectively avoiding repeated failures and promoting the migration of excellent design patterns between different populations. Additionally, VFlow introduces a multi-level verification process, including syntax/static checking, functional simulation, and boundary condition testing, ensuring that the generated Verilog code meets industrial standards in terms of synthesizability and timing.

Experimental results on public benchmarks such as VerilogEval and RTLLM2.0 demonstrate that VFlow achieves an average improvement of approximately \textbf{20\%-30\%} in functional correctness (pass@1) compared to direct invocation or mainstream multi-prompt methods, and significantly approaches or even exceeds human design reference values in area and power consumption metrics. Simultaneously, through optimized agent workflows, \textbf{VFlow enables smaller specialized LLMs to outperform larger general-purpose models} in Verilog generation tasks, significantly reducing inference overhead and costs, exhibiting an excellent return on investment (ROI).

\vspace{0.2cm}
\noindent
\textbf{Our main contributions} are summarized as follows:
\begin{enumerate}
    \item \textbf{MCTS-driven Hardware LLM Workflow Framework:} We propose a pioneering automatic discovery framework specifically designed for hardware design tasks using Monte Carlo Tree Search.
    
    \item \textbf{Domain-specific Verification Operators:} We design hardware-specific verification operators that systematically ensure the syntax correctness, functional validity, and synthesizability of generated Verilog code.
    
    \item \textbf{Multi-objective Optimization Mechanisms:} We construct multi-population cooperation, experience sharing, and fragment migration mechanisms to accelerate and enhance multi-objective search in the hardware design space.
    
    \item \textbf{Model Efficiency Improvements:} We experimentally demonstrate that our method enables smaller specialized models to outperform larger general-purpose models, offering significant cost-effectiveness and practical potential for real-world hardware design tasks.
\end{enumerate}

\section{Background}
\subsection{Hardware Description Languages and Verilog}

HDLs serve as the primary tools for describing digital circuits and systems. Unlike software programming languages that execute sequentially, HDLs describe parallel hardware structures and behaviors. 

Verilog designs center around modules—fundamental building blocks with defined interfaces. These modules contain port declarations, internal signal definitions, behavioral descriptions using \texttt{always} blocks, and data manipulations through \texttt{assign} statements. In complex designs, modules can instantiate other modules, creating hierarchical structures that mirror physical hardware organization.

Verilog code generation presents unique challenges: (1)\textbf{Hardware Semantics} describes concurrent hardware behavior; (2)\textbf{Timing Considerations} require attention to synchronous and asynchronous behaviors; (3)\textbf{Synthesizability} constraints mean not all valid Verilog code can be synthesized into physical hardware; (4)\textbf{Verification Requirements} typically demand simulation-based testbench verification.

\subsection{LLMs for Verilog Code Generation}

In recent years, Verilog code generation based on LLMs has achieved significant progress, demonstrated by rapid improvements in benchmark performance, novel model architectures, and gradual expansion to applications across the entire hardware design workflow. Pinckney et al. \cite{pinckney2025revisiting} enhanced the VerilogEval benchmark by incorporating specification-to-RTL generation tasks and in-context learning (ICL). In this test, GPT-4o achieved a pass rate of approximately 63\%, Llama 3.1 405B reached 58\%, while RTL-Coder (6.7B)—a smaller model specifically trained for hardware design—attained a pass rate of about 34\%. The graph-enhanced model RTL++ significantly improved generation correctness by encoding control flow/data flow graphs in textual form \cite{akyash2025rtl++}. Meanwhile, CodeV-R1 introduced Reinforcement Learning with Verifiable Rewards (RLVR), achieving 68.6\% pass@1 on VerilogEval v2 and 72.9\% on RTLLM v1.1, outperforming previous best models by 12-20 percentage points \cite{zhu2025codev}.

Furthermore, more research focusing on data-oriented approaches and hallucination control for RTL generation has emerged. For example, HaVen \cite{yang2025haven} significantly reduced hallucination rates in Verilog generation through chain-of-thought reasoning. RTLLM2.0 and RTLCoder-Data expanded task scale and example diversity, providing models with more adequate training and evaluation resources \cite{lu2024rtllm,liu2024rtlcoder}.

Notably, LLM applications are no longer limited to code generation. Recent research shows that LLMs are also being used for root cause analysis of simulation/synthesis bugs \cite{qiu2025llmbasedrootcauseanalysis} and demonstrate good results in coverage-driven test stimulus generation (e.g., LLM4DV) \cite{zhang2025llm4dv}. A survey paper on hardware design points out that there are now over 70 LLM papers specifically focused on hardware, covering scenarios such as synthesis, formal verification, and bug detection, with future work urgently needed in domain-specific models, hybrid formal/machine learning systems, and explainability \cite{abdollahi2025hardware}.

These advances indicate that as benchmark systems continue to improve, and model architectures continue to innovate, LLMs have gradually transitioned from natural language and software code generation to the hardware domain, becoming indispensable intelligent components in RTL design workflows. Small to medium-scale hardware-specific models (such as RTL-Coder) demonstrate higher parameter efficiency compared to general-purpose models, further confirming the value of targeted training for hardware design.

\begin{figure*}[htbp]
\centering
\includegraphics[width=0.95\linewidth]{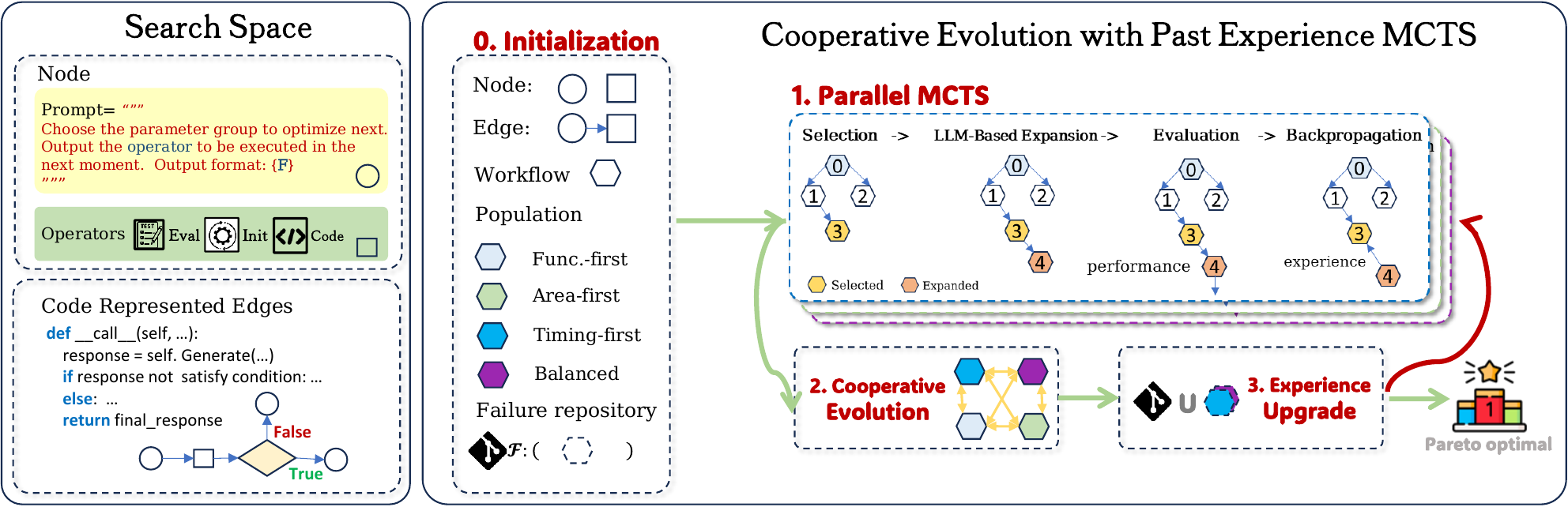}
\caption{Workflow of VFlow.}
\label{fig:vflow}
\vspace{-0.4cm}
\end{figure*}

\subsection{Agentic Workflows}

Agentic workflows have emerged as a powerful paradigm for extending LLM capabilities through structured sequences of invocations with detailed instructions. As Zhang et al. \cite{zhang2024aflow} note, these workflows enable LLMs to tackle complex tasks across diverse domains, from code generation and data analysis to question answering and mathematical reasoning.

Traditionally, agentic workflows relied on manual design, requiring significant human effort to create and optimize. Recent research has focused on automating the discovery and optimization of these workflows, with methods ranging from prompt optimization \cite{khattab2023dspy} to comprehensive workflow automation \cite{hu2024automated}.

A key innovation in this field is reconceptualizing workflow optimization as a search problem over code-represented workflows. In this framework, "LLM-invoking nodes are connected by edges" that define "the logic, dependencies, and flow between these actions" \cite{zhang2024aflow}. This representation transforms the workflow into a vast search space that can be systematically explored using techniques such as Monte Carlo Tree Search.

The automation of agentic workflow discovery represents a promising frontier for AI research. By reformulating workflow design as a search problem, approaches like \textit{AFLOW} and \textit{MaAS} \cite{zhang2025multi} can systematically discover effective workflows without extensive human intervention.

\section{VFlow Framework}
\subsection{Problem Formulation}

We formulate Verilog code generation as a search problem over workflows that transform hardware specifications into correct, efficient code. A Verilog generation workflow $W$ can be represented as a sequence of LLM-invoking nodes:

\begin{equation}
W = (N, E)
\end{equation}
where $N = \{N_1, N_2, ..., N_k\}$ is the set of nodes and $E$ represents the connections between these nodes. Each node $N_i$ is characterized by four key parameters:

\begin{equation}
N_i = (M_i, P_i, \tau_i, F_i)
\end{equation}
where $M_i$ is the language model, $P_i$ is the prompt, $\tau_i$ is the temperature setting, and $F_i$ is the output format. The edges $E$ define the flow of information between nodes, establishing the execution sequence of the workflow.

Given a hardware design task $T$ and an evaluation function $G$ that measures performance, VFlow aims to discover the optimal workflow $W^*$ that produces the highest-quality Verilog code:

\begin{equation}
W^* = \arg\max_{W} G(W, T)
\end{equation}
The search space encompasses all possible workflow configurations, including different combinations of models, prompts, parameters, and connections.

\subsection{Architecture of VFlow}
This paper proposes the CEPE-MCTS (Cooperative Evolution with Past Experience Monte Carlo Tree Search) algorithm, which is a multi-population cooperative search framework designed for Verilog code generation. The algorithm efficiently discovers and optimizes workflows by maintaining multiple specialized search populations to explore different optimization objectives in parallel, while establishing a global failure experience sharing mechanism to avoid repetitive errors.

\subsubsection{Multi-population Cooperative Architecture}
The core concept of CEPE-MCTS is to decompose the traditional single search process into multiple specialized parallel search populations. We define a set of $K$ specialized search populations as $\mathcal{P} = \{P_1, P_2, ..., P_K\}$, where each population $P_i$ focuses on a specific design objective. Each population can be represented as a triplet $P_i = (T_i^{MCTS}, f_i^{obj}, \mathcal{F}_i^{shared})$, where $T_i^{MCTS}$ is the MCTS search tree maintained by the population, $f_i^{obj}$ is the population's specialized objective function, and $\mathcal{F}_i^{shared}$ is the failure experience repository shared with other populations.

To address the multi-objective nature of Verilog code generation, we design four specialized populations: functionality-first population optimizing functional correctness $f_1^{obj}(W) = G_{functional}(W)$, area optimization population minimizing hardware resource consumption $f_2^{obj}(W) = G_{area}(W)$, timing optimization population focusing on meeting clock constraints $f_3^{obj}(W) = G_{timing}(W)$, and balanced population considering all objectives comprehensively $f_4^{obj}(W) = \sum_{j=1}^{3} w_j G_j(W)$. This specialization design enables each population to conduct deep searches within its area of expertise, avoiding the objective conflict issues in traditional multi-objective optimization.

\subsubsection{Fragment Migration Mechanism for Cooperative Evolution}
To achieve knowledge sharing and cooperative optimization among populations, we design an intelligent fragment migration mechanism. The core of this mechanism is to evaluate the fitness of workflow fragments across different populations to determine whether cross-population migration should occur. We define the fitness evaluation function for workflow fragments as follows:

\begin{equation}
    \phi(W_{\text{frag}}, P_j) = \frac{f_j^{obj}(W_{\text{frag}} \oplus W_{\text{base}}) - f_j^{obj}(W_{\text{base}})}{f_j^{obj}(W_{\text{base}})}
\end{equation}

where $W_{\text{frag}}$ represents the candidate workflow fragment for migration, $W_{\text{base}}$ is the base workflow in the target population, and $W_{\text{frag}} \oplus W_{\text{base}}$ denotes the operation of integrating the fragment into the base workflow. The decision rule for fragment migration is based on a fitness threshold: when $\phi(W_{\text{frag}}, P_j) > \theta_{\text{migrate}}$, the algorithm executes the fragment migration operation from the source population to the target population. This performance-based migration strategy ensures that only truly valuable design patterns propagate between populations, avoiding the negative impact of ineffective migrations on search efficiency. The migration process is executed at fixed time intervals, allowing populations to regularly exchange excellent design experiences, forming a dynamic cooperative evolution process.

\subsubsection{Global Failure Experience Sharing Mechanism}
To address complex failure modes in Verilog code generation, we establish a global failure experience sharing mechanism. This mechanism maintains a unified failure experience repository $\mathcal{F}^{global} = \bigcup_{i=1}^{K} \mathcal{F}_i^{local}$, recording failure cases encountered by all populations. Each failure record includes four elements: the structure of the failed workflow $W'_i$, the failure type $e_i$, the severity of the failure $s_i$, and the timestamp when the failure occurred $t_i$. To handle the timeliness of experiences, we introduce a time decay mechanism to dynamically adjust the influence weight of historical failure experiences. The failure risk assessment function based on workflow similarity is defined as:
\begin{equation}
    r_k(W) = \max_{W'_i \in \mathcal{F}_k} \text{sim}(W, W'_i) \cdot s_i \cdot \exp(-\alpha(t_{current} - t_i)
\end{equation}
where $\mathcal{F}_k$ represents the subset of failure cases for a specific failure type $k$, and $\text{sim}(W, W'_i)$ is the workflow structure similarity function, calculated using the Jaccard similarity coefficient: $\text{sim}(W, W') = \frac{|\text{nodes}(W) \cap \text{nodes}(W')|}{|\text{nodes}(W) \cup \text{nodes}(W')|}$. The time decay factor $\exp(-\alpha(t_{current} - t_i))$ ensures that recent failure experiences have higher weights, while outdated failure patterns gradually reduce their influence.

\subsubsection{Cooperative Enhanced Selection Strategy}
Building upon the traditional MCTS selection strategy, we design an enhanced selection strategy that comprehensively considers population-specialized objectives, failure risks, and inter-population diversity. The selection function for each population is defined as follows:

\begin{equation}
\begin{aligned}
UCB_{CEPE}^{(i)}(n) = {} & f_i^{obj}(\bar{W}_n)
+ C \sqrt{\frac{\ln N}{n_n}} \\
& - \lambda \sum_{k} r_k(W_n)
+ \beta \cdot \text{diversity}(W_n, \mathcal{P}_{-i})
\end{aligned}
\end{equation}

The first term $f_i^{obj}(\bar{W}_n)$ represents the average performance of node $n$ under the population $i$'s specialized objective, the second term $C\sqrt{\frac{\ln N}{n_n}}$ is the classical confidence bound term for balancing exploration and exploitation, the third term $-\lambda \sum_{k} r_k(W_n)$ is the failure risk penalty term, and the fourth term $\beta \cdot \text{diversity}(W_n, \mathcal{P}_{-i})$ is the diversity reward term. The diversity function is defined as $\text{diversity}(W_n, \mathcal{P}_{-i}) = \frac{1}{K-1} \sum_{j \neq i} \min_{W' \in P_j} (1 - \text{sim}(W_n, W'))$, used to encourage different populations to explore different search directions, avoiding overlap in the search space. Parameters $\lambda$ and $\beta$ control the intensity of failure avoidance and diversity incentivization, respectively, and can be adjusted according to the specific characteristics of the problem.

\subsubsection{Algorithm Execution Flow}
The entire process of the CEPE-MCTS algorithm is divided into three core phases: parallel population search, cooperative evolution, and global experience update. In the parallel search phase, all populations simultaneously execute the standard steps of MCTS, with the key innovation being the failure risk pre-check during the expansion phase: when the similarity risk between a new workflow and historical failure cases exceeds the threshold $\theta_{risk}$, the candidate is rejected in advance, avoiding repetitive errors. The cooperative evolution phase periodically performs cross-population knowledge migration, evaluating fragment value through the fitness function $\phi(W_{frag}, P_j)$, and executing migration only when the fitness gain exceeds $\theta_{migrate}$, ensuring the effectiveness of knowledge propagation. The global experience update phase maintains a unified failure experience repository, applies the time decay mechanism, and updates the Pareto frontier, ultimately outputting multiple Pareto-optimal workflow solution sets.

\subsection{Domain-Specific Considerations}

\subsubsection{Multi-Level Simulation Verification}
VFlow implements a progressive verification strategy, achieving a balance between evaluation depth and computational efficiency. This strategy comprises three verification levels: the syntax and static analysis level performs preliminary filtering based on Verilog syntax correctness, the functional simulation level uses test vectors to cover key paths for behavioral verification, and the boundary case verification level executes comprehensive testing including boundary conditions. The verification functions are defined respectively as a binary syntax check function $V_1(W,T) \in \{0,1\}$, a test pass rate function $V_2(W,T) = \frac{\text{number of passed tests}}{|\text{total number of tests}|}$, and the strictest all-pass requirement $V_3(W,T) = \min_t \text{test result}(t)$. This multi-level approach enables the algorithm to quickly eliminate non-functional designs, avoiding wasting computational resources on invalid candidates.

\subsubsection{Hierarchical Design Pattern Support}

\definecolor{bestarea}{rgb}{0.85, 0.95, 0.85}     
\definecolor{bestpower}{rgb}{0.85, 0.85, 0.95}    
\definecolor{besttiming}{rgb}{0.95, 0.95, 0.85}   
\definecolor{bestmultiple}{rgb}{0.95, 0.85, 0.95} 

\begin{table*}[!t]
\centering
\renewcommand\arraystretch{1.2}
\caption{Detailed Performance Metrics with Best Values Highlighted}
\begin{adjustbox}{max width=\textwidth}
\begin{tabular}{@{}l*{9}{c}@{}}
\toprule
\multirow{3}{*}{\textbf{Method}} & \multicolumn{9}{c}{\textbf{Hardware Modules}} \\
\cmidrule{2-10}
& \textbf{mux} & \textbf{serial2parallel} & \textbf{edge\_detect} & \textbf{adder\_16bit} & \textbf{multi\_16bit} & \textbf{JC\_counter} & \textbf{ALU} & \textbf{PE} & \textbf{parallel2serial} \\
\cmidrule{2-10}
& \multicolumn{9}{c}{\footnotesize{\textcolor{green!60!black}{Area ($\mu m^2$)} / \textcolor{blue!60!black}{Power ($\mu W$)} / \textcolor{orange!60!black}{Timing (ns)},  best values highlighted with background colors}} \\
\midrule
Designer Ref. & 
\cellcolor{bestmultiple}\textbf{68}/\textbf{6.5}/-0.08 & 
\cellcolor{bestmultiple}\textbf{168}/\textbf{16K}/-0.30 & 
\cellcolor{bestmultiple}\textbf{19}/\textbf{2.6K}/-0.14 & 
128/\textbf{68}/-1.21 & 
\cellcolor{bestmultiple}\textbf{749}/\textbf{75K}/-0.91 & 
\cellcolor{bestmultiple}\textbf{380}/\textbf{45K}/-0.13 &
\cellcolor{bestmultiple}\textbf{2.4K}/\textbf{36K}/-1.03 & 
\cellcolor{bestmultiple}\textbf{2.4K}/\textbf{36K}/-1.03 & 
\cellcolor{bestarea}\textbf{35}/6.2K/-0.21 \\

IO & 
\cellcolor{besttiming}75/8/\textbf{-0.07} & 
\cellcolor{besttiming}200/18K/\textbf{-0.28} & 
\cellcolor{besttiming}23/3.3K/\textbf{-0.12} &
\cellcolor{besttiming}157/91/\textbf{-0.33} & 
\cellcolor{besttiming}930/95K/\textbf{-0.50} & 
\cellcolor{besttiming}410/48K/\textbf{-0.10} &
\cellcolor{besttiming}2.6K/39K/\textbf{-0.90} & 
\cellcolor{besttiming}2.6K/39K/\textbf{-0.90} & 
\cellcolor{besttiming}42/6.5K/\textbf{-0.20} \\

PromptV &
70/7.2/-0.08 & 180/16.5K/-0.29 & 20/2.9K/-0.13 &
135/80/-0.37 & 840/86K/-0.58 & 390/46K/-0.12 &
2.48K/36.5K/-0.97 & 2.48K/36.5K/-0.97 & 36/6.1K/-0.21 \\

AFlow &
69/7/-0.08 & 175/16.2K/-0.29 & \cellcolor{bestarea}\textbf{19}/2.8K/-0.14 &
130/76/-0.38 & 800/83K/-0.60 & 385/\cellcolor{bestpower}\textbf{45K}/-0.13 &
2.46K/36.2K/-1.00 & 2.46K/36.2K/-1.00 & \cellcolor{bestmultiple}\textbf{35}/\textbf{6.0K}/-0.21 \\

\textbf{VFlow} &
\cellcolor{bestarea}\textbf{68}/6.6/-0.09 & 
169/\cellcolor{bestpower}\textbf{16K}/-0.31 & 
\cellcolor{bestmultiple}\textbf{19}/\textbf{2.6K}/-0.14 &
\cellcolor{bestarea}\textbf{126}/69/-1.20 & 
750/\cellcolor{bestpower}\textbf{75K}/-0.90 & 
\cellcolor{bestmultiple}\textbf{380}/\textbf{45K}/-0.13 &
2.43K/\cellcolor{bestpower}\textbf{36K}/-1.02 & 
2.43K/\cellcolor{bestpower}\textbf{36K}/-1.02 & 
\cellcolor{bestmultiple}\textbf{35}/\textbf{6.0K}/-0.22 \\
\bottomrule
\end{tabular}
\end{adjustbox}

\label{tab:detailed_performance_metrics}
\end{table*}

\begin{table}[!t]
\vspace{-0.1cm}
\centering
\renewcommand\arraystretch{1.2}
\setlength{\abovecaptionskip}{0cm}
\caption{Performance Comparison on VerilogEval Benchmark}
\begin{tabular}{@{}l|cc|cc@{}}
\toprule
\multirow{2}{*}{\textbf{Method}} & 
\multicolumn{2}{c|}{\textbf{VerilogEval-machine}} & 
\multicolumn{2}{c}{\textbf{VerilogEval-human}}\\
\cmidrule{2-3} \cmidrule{4-5}
& \textbf{pass@1} & \textbf{pass@5} & \textbf{pass@1} & \textbf{pass@5} \\
\midrule
IO & 56.7\% & 79.1\% & 46.6\% & 60.8\% \\
\rowcolor{gray!8}
COT & 60.5\% & 81.2\% & 48.4\% & 62.9\% \\
Self-Consistency CoT & 62.3\% & 83.8\% & 51.5\% & 64.3\% \\
\rowcolor{gray!8}
MultiPersona Debate & 64.1\% & 84.7\% & 55.2\% & 66.7\% \\
Self-Refine & 61.9\% & 82.1\% & 59.8\% & 73.4\% \\
\rowcolor{gray!8}
PromptV & 75.7\% & 87.9\% & 79.3\% & 86.2\% \\
AFlow & 80.1\% & 91.2\% & 80.6\% & 88.1\% \\
\rowcolor{blue!12}
\textbf{VFlow} & \textbf{84.3\%} & \textbf{96.5\%} & \textbf{82.8\%} & \textbf{91.5\%} \\
\midrule
\rowcolor{green!10}
\textbf{Improvement over IO} & \textbf{+27.6\%} & \textbf{+17.4\%} & \textbf{+36.1\%} & \textbf{+30.7\%} \\
\bottomrule
\end{tabular}
\label{tab:performance_compact}
\vspace{-0.4cm}
\end{table}

Hardware designs naturally decompose into hierarchical modules. VFlow incorporates this domain knowledge through specialized patterns that encourage modularity and reuse:

\begin{multline}
    M(W) = \psi_1 \cdot \frac{|\text{modules}|}{|\text{total\_logic}|}
    + \psi_2 \cdot \frac{|\text{reused\_modules}|}{|\text{modules}|} \\
    - \psi_3 \cdot \max(0, D_{\text{hier}} - D_{\text{max}})
\end{multline}

where $D_{\text{hier}}$ represents the hierarchical depth of the design and $D_{\text{max}}$ is the maximum recommended depth. This metric rewards appropriate modularization while penalizing excessive hierarchy that might impact synthesis results.

\subsubsection{Verilog-Specific Operator Design}
To better support the special requirements of Verilog code generation, CEPE-MCTS extends the original set of generic operators by designing and implementing a series of domain-specific operators targeted at hardware description languages. These specialized operators include the Syntax Validator operator for enforcing Verilog language rules and constraints, the Simulation Executor operator for interfacing with simulators such as Icarus Verilog for functional verification, the Waveform Analyzer operator for evaluating the consistency between simulation results and expected behavior, the Circuit Optimizer operator for optimizing generated code based on area, power consumption, or timing constraints, and the Hierarchical Composer operator for assembling modular components into complete designs. The introduction of these specialized operators enables CEPE-MCTS to more precisely address various technical challenges in the Verilog code generation process.

\section{Experiments}
\subsection{Experimental Setup}

\textbf{Datasets.} We evaluate VFlow’s performance across a diverse range of hardware design tasks drawn from the VerilogEval \cite{liu2023verilogeval} benchmark suite, which consists of 156 problems spanning simple combinational circuits to complex finite state machines. Additionally, we select 9 representative modules from the RTLLM2.0 \cite{lu2024rtllm} dataset, covering arithmetic operators, memory structures, and control logic, to further assess VFlow’s capability on more realistic and medium-scale RTL designs.

\textbf{Baselines.} For model selection, we employ four LLMs spanning different sizes and capabilities: GPT-4o-mini, DeepSeek-V3 \cite{liu2024deepseek}, DeepSeek-R1 \cite{guo2025deepseek}, Claude-3.7-Sonnet, and GPT-4o. All models are accessed via their respective APIs with consistent temperature settings. To ensure fair comparison, we implement seven baseline methods: IO (direct LLM invocation), Chain-of-Thought \cite{wei2022chain}, Self-Consistency CoT \cite{wang2022self}, MultiPersona Debate \cite{wang2023unleashing}, Self-Refine \cite{madaan2023self}, PromptV \cite{mi2024promptv} and AFlow \cite{zhang2024aflow}. Since VFLow autonomously selects the appropriate LLM, these baselines all use the most powerful overall performer, Claude-3.7-Sonnet, to ensure fairness.

\textbf{Implementation Details.} For VFlow's MCTS optimization, we set the maximum iteration rounds to 20, with early stopping triggered after 5 rounds without improvement. We instantiate the simulator interface with Icarus Verilog for functional verification and Yosys for synthesis metrics, allowing for comprehensive evaluation of both behavioral correctness and circuit quality. For simple tasks, we only consider the trade-off between pass rate and API cost.

\subsection{Experimental Results and Analysis}

\subsubsection{Overall Performance Comparison}

\begin{figure*}[htbp]
\centering
\includegraphics[width=0.95\linewidth]{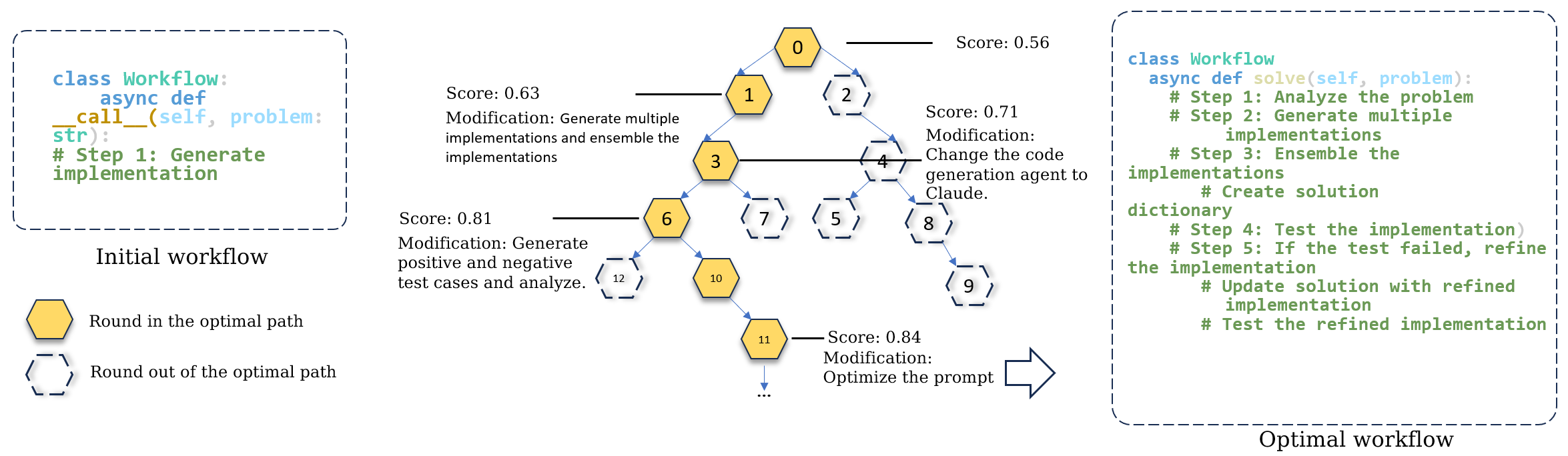}
\caption{VFlow's MCTS-Based Workflow Evolution for Verilog Code Generation.}
\label{fig:optimized}
\vspace{-0.4cm}
\end{figure*}

\begin{figure}[htbp]
\centering
\includegraphics[width=0.9\linewidth]{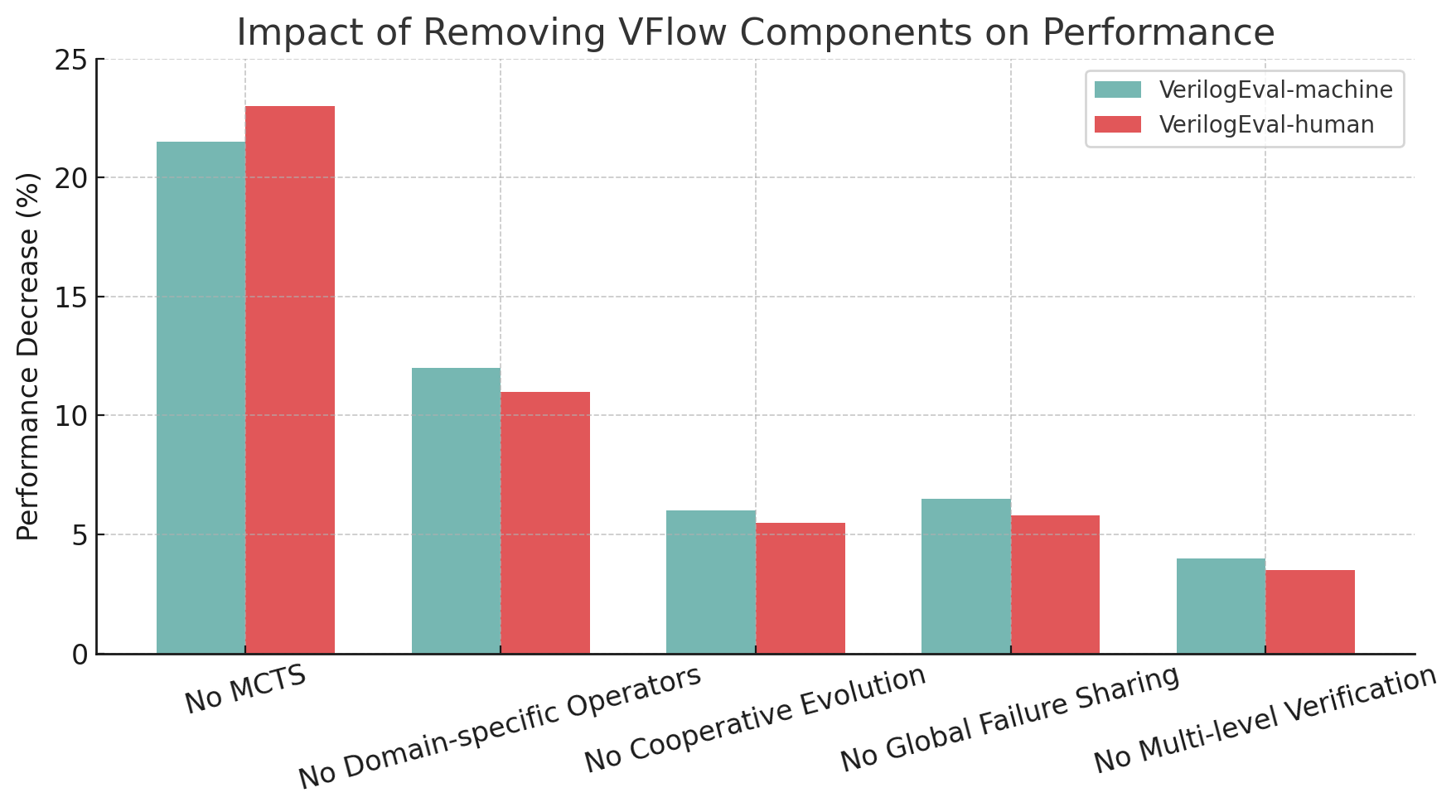}
\caption{Ablation Studies on VFlow Components (pass@1 performance).}
\label{fig:ablation}
\vspace{-0.4cm}
\end{figure}

Table~\ref{tab:performance_compact} summarizes the overall success rates of different methods on the VerilogEval benchmark. Compared to direct invocation (IO) and prior prompting strategies, our proposed VFlow framework achieves the highest pass@1, pass@5, and pass@10 across both evaluations, with an average improvement of +19.6\% (machine) and +30.9\% (human) over IO. This clearly demonstrates VFlow’s effectiveness in consistently generating functionally correct Verilog code across a wide range of tasks.

Table~\ref{tab:detailed_performance_metrics} provides a fine-grained comparison on nine representative hardware modules, reporting area, power, and timing metrics. We observe that VFlow either matches or surpasses the best baselines in most cases, achieving minimal area and power in several modules (e.g., \textit{mux}, \textit{adder\_16bit}, \textit{serial2parallel}), while also maintaining competitive timing slack close to the designer reference. This performance gain can be attributed to VFlow’s cooperative multi-population search framework, which allows different specialized populations to deeply explore distinct objectives (such as minimizing area or power) without conflicting trade-offs. Moreover, the fragment migration and global failure sharing mechanisms enable VFlow to effectively propagate promising design patterns and systematically avoid recurring inefficiencies across different search threads. These results demonstrate that beyond generating functionally correct code, VFlow consistently produces RTL designs with superior implementation quality, striking a robust balance among hardware resource efficiency, power consumption, and timing closure.

\begin{table}[!t]
\vspace{-0.1cm}
\centering
\renewcommand\arraystretch{1.2}
\setlength{\abovecaptionskip}{0cm}
\caption{Cost-Performance Analysis: VFlow vs. Baseline Methods}
\begin{adjustbox}{max width=\textwidth}
\begin{tabular}{@{}l|cc|cc|c@{}}
\toprule
\multirow{2}{*}{\textbf{Model}} &
\multicolumn{2}{c|}{\textbf{Performance}} &
\multicolumn{2}{c|}{\textbf{Efficiency}} &
\multirow{2}{*}{\textbf{ROI}}\\
\cmidrule{2-3} \cmidrule{4-5}
& \textbf{pass@1} & \textbf{Relative} & \textbf{Tokens} & \textbf{Cost} & \\
\midrule
GPT-4o (IO) & 57.0\% & 100\% & 1.0× & 1.00× & 1.00× \\
\rowcolor{blue!12}
\textbf{DeepSeek-V3 (VFlow)} & \textbf{80.5\%} & \textbf{+41.2\%} & \textbf{0.4×} & \textbf{0.13×} & \textbf{10.9×}\\
\rowcolor{green!8}
\textbf{GPT-4o-mini (VFlow)} & \textbf{76.7\%} & \textbf{+34.7\%} & \textbf{0.7×} & \textbf{0.42×} & \textbf{3.2×} \\
\bottomrule
\end{tabular}
\end{adjustbox}
\label{tab:cost-performance-detailed}
\vspace{-0.4cm}
\end{table}

\subsubsection{Cost-Performance Analysis}
Table~\ref{tab:cost-performance-detailed} presents a comparative cost-performance analysis, highlighting both effectiveness and efficiency across different configurations. We observe that VFlow, when applied on DeepSeek-V3, achieves a substantial pass@1 improvement of +41.2\% over the GPT-4o baseline, while simultaneously reducing token consumption and cost to just 0.4× and 0.13× respectively, yielding an impressive 10.9× return on investment (ROI). Similarly, integrating VFlow with GPT-4o-mini delivers a notable 34.7\% higher pass@1 at less than half the cost. These results demonstrate that VFlow not only enhances design quality, but also significantly lowers the generation overhead, making it a highly cost-effective solution for large-scale RTL code generation.

\subsubsection{Ablation Studies}
To evaluate the contribution of each component in VFlow, we conduct a series of ablation studies by selectively removing key modules, including MCTS optimization, domain-specific operators, cooperative evolution, global failure experience sharing, and multi-level verification. As shown in Fig.~\ref{fig:ablation}, removing MCTS results in the most significant performance drop (over 20\%), followed by the absence of domain-specific operators (around 10\%). Eliminating cooperative evolution or global failure sharing leads to moderate degradations (approximately 5--7\%), while skipping multi-level verification causes a minor decline (about 3--4\%). These results confirm that each component synergistically enhances VFlow’s effectiveness in generating high-quality Verilog code.

\subsubsection{Discovered Workflow Analysis}
The MCTS-based optimization in VFlow demonstrates a significant evolutionary progression from a simplistic implementation to a sophisticated multi-step workflow as shown in Figure \ref{fig:optimized}. Beginning with a basic single-step approach (score: 0.56), the system progressively discovers more effective strategies through key modifications: generating multiple implementations (round 1, 0.63), incorporating Claude as the code generation agent (round 4, 0.71), introducing comprehensive test cases (round 6, 0.81), and ultimately optimizing prompts (round 11, 0.84). The discovered optimal workflow establishes a robust five-step process of problem analysis, multiple implementation generation, ensemble integration, comprehensive testing, and targeted refinement - effectively transforming Verilog code generation from a straightforward task into a verification-integrated process with continuous improvement capabilities.



\section{Conclusion}
We presented VFlow, an automated framework that discovers optimized agentic workflows for Verilog code generation. By integrating MCTS-based search, domain-specific operators, and multi-level verification, VFlow achieves higher code quality and efficiency compared to existing methods. Our experiments demonstrate its strong potential for advancing hardware design automation.

\newpage
\bibliographystyle{IEEEtran}
\bibliography{ref}

\end{document}